\documentclass[doublecol]{epl2} 

\usepackage{amsfonts}
\usepackage{amsmath}
\usepackage{amssymb}
\usepackage{graphicx}
\usepackage{bm}

\renewcommand{\d}{\text{d}}

\title{Probing nanoscale deformations of a fluctuating interface}
\shorttitle{Probing nanoscale deformations of a fluctuating interface} 

\author{T. Bickel}
\shortauthor{T. Bickel}

\institute{ 
Laboratoire Ondes et Mati\`ere d'Aquitaine,  Universit\'{e} de Bordeaux \& CNRS,  33405 Talence, France
}
\pacs{68.05.-n}{Liquid-liquid interfaces}
\pacs{68.03.Cd}{Surface tension and related phenomena} 
\pacs{05.70.Np}{Interface and surface thermodynamics}

\abstract{
We consider the contribution of thermal capillary waves to the interaction between a fluid-fluid interface and a nearby nanoparticle.
Fluctuations are described thanks to an effective  interaction potential which is derived using 
the renormalization group.  
The general theory is then applied to a spherical particle interacting with the interface through van der Waals forces.
Surprisingly enough, we find that fluctuations contribute significantly to the deformation profile. Our study therefore reveals that
thermal fluctuations cannot be ignored when probing nanoscale deformations of a soft interface.
}

\begin{document}

\maketitle

\section{Introduction}

With the miniaturization of fluidic devices, it is now possible to study simple and complex fluids at the scale of the nanometer.
As the size of the system decreases, 
confinement as well as the importance taken by surface effects are expected to lead to
novel transport properties~\cite{Boc10,Sir13}. 
Accordingly, exploiting the possibilities of nanofluidics requires a fine knowledge  of liquids and liquid interfaces at very small scales.
In this context, nanoscale measurements of liquid-surface properties  have become increasingly popular.
For instance, the contribution of individual surface defects to contact angle hysteresis has been evidenced using  atomic force microscopy (AFM) with a carbon nanotube probe~\cite{Del11}. Local rheology measurements have been performed using a hanging-fiber AFM~\cite{Dev13}, and
nanoscale deformations of an interface in response to the  interactions with an AFM  tip
have been characterized recently~\cite{Led12a}. Non-contact manipulation of  liquid interfaces 
has also been achieved using magnetic beads~\cite{Tsa13} or knife-edge electric field tweezers techniques~\cite{Shi11,Shi13}, whereas
other experiments with near-critical fluids have explored interfacial deformations by a laser beam~\cite{Cas01}.

Despite continuous progress, some fundamental issues regarding the properties of a liquid interface at the nanoscale remains unsolved.
The surface of a liquid is rather difficult to probe since the interaction with the measuring device is expected to induce strong perturbations of the interface~\cite{Rap96}. 
A thorough description of the probe-interface interaction is therefore required in order to get conclusive information regarding liquid parameters such as surface tension or  viscosity at very small  scales.
In recent years, several groups have developed theoretical approaches to describe the interaction of an interface  with a nanoscopic probe~\cite{Liu05,Led12b,Led13,Qui13}. The resulting deformation  is obtained as the minimum of some elastic energy, but thermally activated fluctuations have been systematically overlooked so far.
However,  interfacial fluctuations  range from a few angstroms to a few nanometers~\cite{Lan92} and are therefore expected to become relevant 
when the size of the probe reaches the nanometer scale. 
This assertion is also valid for larger probes if the fluids are close to a critical point,
in which case fluctuations can lie in the micrometer range~\cite{Aar04}.

The aim of this Letter is thus to study theoretically  the effect of thermal capillary waves on interfacial deformation. 
We follow here a linear renormalization group (RG) scheme that is commonly used in the context of wetting.
Indeed,  the critical exponents that characterize
the adsorption transition of a liquid film on a solid substrate
are affected by thermal fluctuations~\cite{Fis04,comment}. 
To account for the latter,
an effective potential  can be derived by tracing out small wavelength fluctuations~\cite{Bre83,Fis85}.
This  renormalized potential is then expressed as a convolution of the bare potential with the fluctuation probability distribution function~\cite{Ind13}.
The very same idea is applied in this work in order to obtain an effective probe-interface potential.
The paper is organized as follows. 
We first derive the shape equation for a given interaction potential. The effect of interfacial fluctuations
is discussed next, and the general theory is then applied to  van der Waals forces. The issue of  fluctuations of the probe itself is commented in the last section.

\begin{figure}
\onefigure[width=8cm]{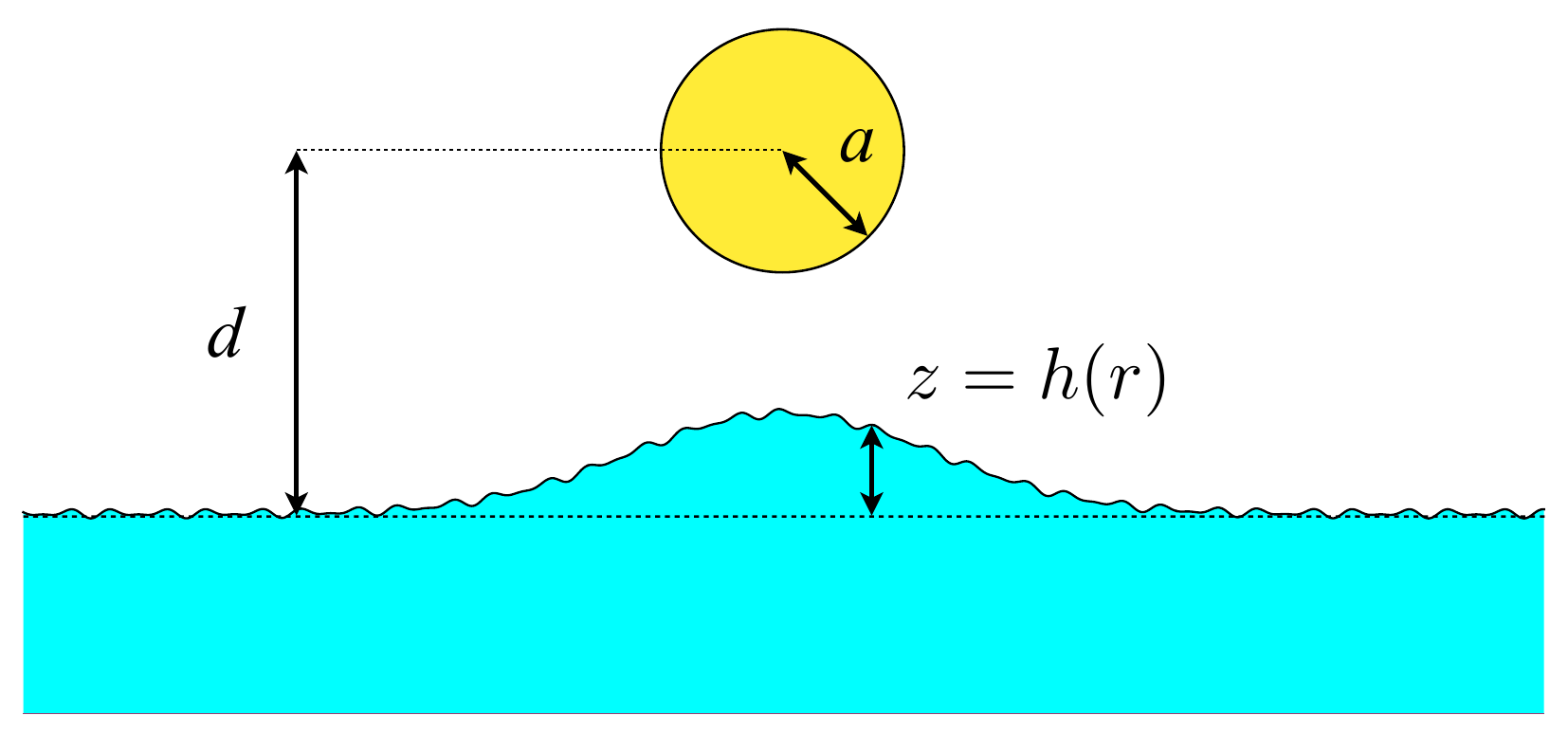}
\caption{Sketch of a spherical probe deforming a fluid-fluid interface. When the size of the probe reaches the nanometer scale,
the interface cannot be assumed to be smooth anymore and
thermally activated  fluctuations are expected to contribute to the deformation profile.}
\label{sketch}
\end{figure}

\section{Euler-Lagrange equation}

We consider a fluid-fluid  interface  which is interacting with a nanoprobe,
as schematically drawn in fig.~\ref{sketch}. 
We limit the discussion to the non-wetting case of a probe that lies at a given height above the interface,
in order to avoid possible issues related to capillary contact~\cite{Led12a}.
If the probe were absent, the interface would be flat and coincide with 
the horizontal plane $z=0$.
Interaction with the  probe 
provokes a local deformation of the interface 
described by  a smooth function $z=h(\mathbf{r})$, with $\mathbf{r}=(x,y)$.
The equilibrium profile is then obtained as  the minimum of the total Hamiltonian~$\mathcal{H}$,
the latter being the sum of two contributions: $\mathcal{H} =  \mathcal{H}_{0} + \mathcal{H}_{I} $.
In the small-gradient approximation $\vert \bm{\nabla} h \vert \ll 1$, the capillary-wave Hamiltonian reads
\begin{equation}
\mathcal{H}_{0} =  \frac{\sigma}{2} \int \d^2 \mathbf{r} \left[  \left( \bm{\nabla} h \right)^2 + l_c^{-2} h^2 \right] \ ,
\label{hcw}
\end{equation}
with $\sigma$ the surface tension and $l_c = \left(\sigma/\Delta \rho g\right)^{1/2}$ the capillary length, $\Delta \rho$ being the mass density difference between the two fluids  and $g$ the gravitational acceleration.
The second term describes the interaction between the probe and the interface, and can be written as 
\begin{equation}
\mathcal{H}_{I} =   \int \d^2 \mathbf{r} V(\mathbf{r},h ) \ .
\label{hint}
\end{equation}
The  potential $V(\mathbf{r},h)$, whose explicit form will be specified later, is a function of both the position 
and  the local shape of the interface.
It is assumed to be radially symmetric
and to vanish beyond a typical  distance $a$ ---  for instance, 
the radius of an AFM tip~\cite{Led12a} or the waist of a laser beam~\cite{Cas01}.

Minimizing the total energy functional $\delta \mathcal{H}/\delta h $=0 then leads to the
generalized Young-Laplace equation~\cite{Liu05}
\begin{equation}
 \nabla^2 h - l_c^{-2} h = - \frac{1}{\sigma} \Pi(\mathbf{r}, h )  \ ,
\label{younglapl}
\end{equation}
with $\Pi = - \partial V / \partial z $ the disjoining pressure~\cite{Bar03}. This equation has to be solved with the  
condition that the profile is flat far away from the probe.

We  make the further assumption that the capillary length is  much larger than the size $a$ of the probe
($l_c$ typically lies in the millimeter range).
Still, we need to keep it finite in order to enforce the condition $h(r)\to 0$ when $r \to \infty$~\cite{Qui13}.
Eq.~(\ref{younglapl}) can  then be solved using the method of matched asymptotic~\cite{Liu05,Qui13}.
For $r\gg l_c$, eq.~(\ref{younglapl}) reduces to $ \nabla^2 h_{out} = l_c^{-2} h_{out}$ and the outer solution reads $h_{out}(r) = \alpha  \text{K}_0 (r/l_c)$. Here, $\text{K}_0$ is the modified Bessel function of second kind and $\alpha$ an unknown constant. In the opposite limit $r \ll l_c$, gravitational effects can be neglected and the inner solution
$h_{in}$ follows the equation $\sigma \nabla^2 h_{in}  = -  \Pi(\mathbf{r}, h_{in} ) $. The matched asymptotic method then requires that
\begin{equation}
\lim_{r \to 0} h_{out} (\mathbf{r}) = \lim_{r\to \infty} h_{in}(\mathbf{r}) \ .
\label{matched}
\end{equation}
The approximate solution is finally obtained by adding
 the inner and outer approximations and subtracting their overlapping value, which would otherwise be counted twice.

\section{Thermal fluctuations}

Due to the random motion of the molecules, a liquid interface  is by essence a fluctuating object.
In the absence of interaction,  
the statistical properties of the free interface are set by the capillary-wave Hamiltonian~$\mathcal{H}_0$.  In particular, the probability distribution $\mathcal{P}(h)$ of height fluctuations reads~\cite{Bar03}
\begin{equation}
\mathcal{P}(h) = \frac{1}{\sqrt{2 \pi} \xi_{\perp}  } e^{-h^2/2 \xi_{\perp}^2} \ .
\end{equation}
The width $\xi_{\perp}$  of the distribution  corresponds to
 the mean-square displacement (MSD)  of the free interface
\begin{equation}
\xi_{\perp}^2 = \langle h^2(\mathbf{r}) \rangle_0 =  \frac{k_BT}{2 \pi \sigma} \ln \left( \frac{l_c}{b} \right) \ ,
\label{xi}
\end{equation}
with  $b$  a microscopic cut-off. 
The MSD typically ranges from a few angstroms to a few nanometers  for usual liquid interfaces~\cite{Lan92}, but can be as large as a few microns for near-critical fluids~\cite{Aar04}.

When considering the interaction with an external probe, one has to account for the roughness of the interface that  appears ``fuzzy'' at the scale of the probe ---
see fig.~\ref{sketch}.
In order to anticipate whether 
fluctuations significantly affect the shape of the interface, we define the dimensionless parameter 
$\varepsilon = \xi_{\perp}/a$ (with $a$ is the size of the probe).
Consider for instance the AFM experiment described in~\cite{Led12a}: given the tip radius $a \approx 10$~nm and $\xi_{\perp} \approx 1$~nm, one gets
$\varepsilon \approx 0.1$ so that fluctuations are expected to be relevant.
On the other hand in the experiment with millimeter-size magnetic beads~\cite{Tsa13}, thermal fluctuations can safely be neglected since  $\varepsilon \approx 10^{-6}$.

From a theoretical viewpoint, the appropriate formalism to describe thermal fluctuations at a given length scale is the renormalization group (RG). 
We follow here a linear functional RG scheme that has been developed to describe the wetting transition~\cite{Bre83,Fis85}.
This approach can easily adapted to our geometry even though the potential depends explicitly on the position. 
Starting form the \textit{bare} interaction potential  $V_0(\mathbf{r},h)$, 
thermal fluctuations are traced out through momentum-shell integration (see for instance ref.~\cite{Fis04} for technical details).  
The  resulting RG flow equation can be integrated explicitly, yielding to the \textit{renormalized} potential~\cite{Fis85,Ind13}
\begin{align}
V(\mathbf{r},h)& = \int_{-\infty}^{\infty} \d h' V^{(0)}(\mathbf{r},h')\mathcal{P}(h-h') \nonumber \\
&\doteq V^{(0)}(\mathbf{r},h)*\mathcal{P}(h) \ .
\end{align}
The renormalized potential  is thus obtained as a convolution of the original potential with the probability distribution of height fluctuations~\cite{Ind13}.
The point is that $V(\mathbf{r},h)$ now includes information regarding the roughness of the fluctuating interface. 

The consequences on the deformation profile can  deduced in a straightforward way. 
Let us denote $h^{(0)}$ the solution of the shape eq.~(\ref{younglapl}) in the absence of fluctuations,
\textit{i.e.} with the bare disjoining pressure $\Pi^{(0)} = - \partial V^{(0)}/\partial z$.
The renormalized profile $h=h^{(0)}+ \delta h$ is then solution of the same eq.~(\ref{younglapl}) but with 
the renormalized disjoining pressure $\Pi= -\partial V / \partial z = \Pi^{(0)} * \mathcal{P} $.
Considering the contribution of the fluctuations as a perturbation, we evaluate the correction $\delta h$ at lowest order
in $\varepsilon $.  We first note that, when $\varepsilon \ll 1$, 
the probability distribution $\mathcal{P}$ is narrowly centered around $z=0$. The disjoining pressure is then given by 
\begin{align}
\Pi \left( \mathbf{r},h \right) =  \Pi^{(0)} \left( \mathbf{r},h \right) + \varepsilon^2 \frac{a^2}{2} \frac{\partial^2  \Pi^{(0)}}{\partial z^2}\left( \mathbf{r},h \right) + \ldots   \ ,
\label{renormpi}
\end{align}
so that the  correction to the deformation profile can  be written as  $\delta h =  \varepsilon^2h^{(1)}(r) + o(\varepsilon^2) $.
Since $\varepsilon^2 \propto k_BT$, we find that the lowest-order correction  scales linearly with temperature.

\section{Van der Waals forces}

\begin{figure}
\onefigure[width=8cm]{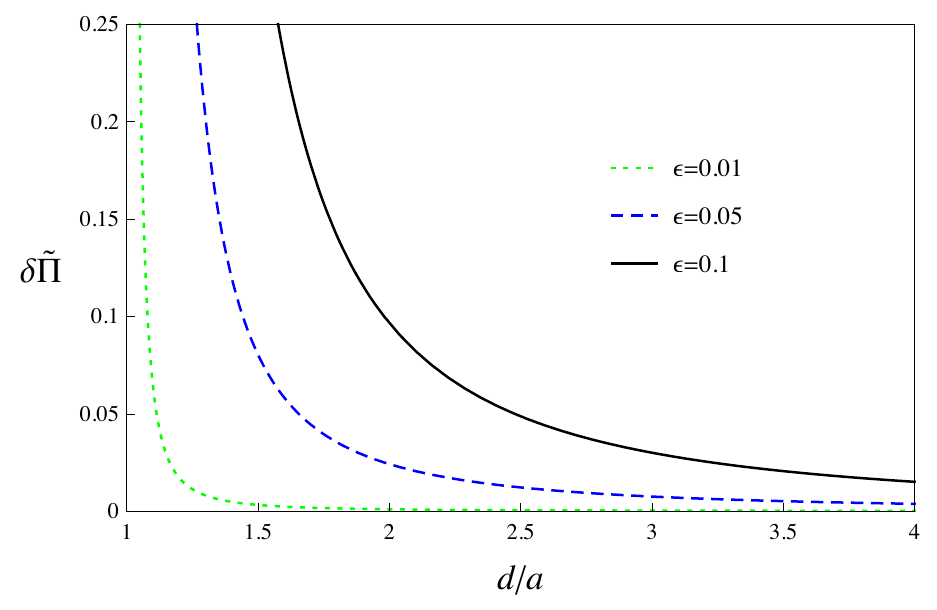}
\caption{Relative increase of the renormalized disjoining pressure $\delta \tilde{\Pi}= (\Pi - \Pi^{(0)}) /\Pi^{(0)}$ evaluated at $\mathbf{r}= \mathbf{0}$ and $z=0$, as a function of the probe-interface distance.}
\label{figpi}
\end{figure}

In the theory developed so far,  no specific assumption  has been made regarding
the interaction potential. The situation that we now discuss is that of a nanoscopic probe interacting with an interface through van der Waals forces.
It is assumed for the sake of simplicity that the probe is a sphere of radius~$a$. The center of the sphere is held at a fixed height $d>a$ above the reference plane --- see fig.~\ref{sketch}.
The attractive force exerted by the sphere over the interface can be
obtained from Hamaker theory~\cite{Isr11}
\begin{equation}
\Pi^{(0)} (\mathbf{r},h)= \frac{ 4 A a^3}{3 \pi}  [ (d-h)^2+r^2-a^2  ]^{-3}   \ ,
\end{equation}
with $A$ the Hamaker constant. The renormalized disjoining pressure $\Pi = \Pi^{(0)} + \varepsilon^2 \Pi^{(1)} + \ldots $ is then deduced from eq.~(\ref{renormpi}) and we get 
\begin{equation}
\Pi^{(1)} (\mathbf{r},h)= \frac{ 4 A a^5}{\pi}  \frac{7(d-h)^2-(r^2-a^2)}{\left[ (d-h)^2+r^2-a^2  \right]^{5}}   \ .
\end{equation}
We plot in fig.~\ref{figpi}
the relative increase $\delta \tilde{\Pi} = \big( \Pi - \Pi^{(0)}  \big) /\Pi^{(0)}$ 
evaluated at $\mathbf{r}= \mathbf{0}$ and $z=0$ (\textit{i.e.}, where the pressure is maximum),
as a function of the particle-interface distance~$d$.
It vanishes as $\delta \tilde{\Pi} (\mathbf{0},0) \sim d^{-2}$ in the limit $d \gg a$, whereas it grows  rapidly when the particle gets closer and closer to the interface. For particle-interface distances that are a few times the particle radius, 
the correction can easily reach 5 or 10~\% of the total pressure.
Note also that, for a given particle-interface distance, the force acting on the interface depends strongly on $\varepsilon$.

Before solving the shape equation, it is convenient to express the relevant energies in terms of dimensionless constants. 
The ratio of van der Waals to surface forces defines the Hamaker number $\mathcal{A}=4A/(3\pi a^2 \sigma)$, whereas
the balance of gravitational to surfaces forces defines  the Bond number $\mathcal{B}=(\Delta \rho g a^2/\sigma)^{1/2}=a/l_c$.
Taking $A\approx 4\times10^{-20}$~J and  $\sigma \approx 10^{-1}$~N/m,  we find that $\mathcal{A} \approx 10^{-3}$ in a typical AFM experiment with $a\approx10$~nm~\cite{Led12a}. We can therefore assume that $\mathcal{A}\ll 1$. This amounts to
evaluate the renormalized disjoining pressure~(\ref{renormpi}) at  $h=0$, and, consequently, the differential eq.~(\ref{younglapl}) becomes linear~\cite{Qui13}. On the other hand the Bond number is also very small
for usual interfaces, at least far from a critical point:  one gets for instance  $\mathcal{B} \approx 3 \times 10^{-6}$ for $a\approx10$~nm and $l_c\approx3$~mm. 
Still this parameter has to be kept finite for technical reasons  in order to enforce the relaxation to the flat shape far away from the particle. 

\begin{figure}
\onefigure[width=8cm]{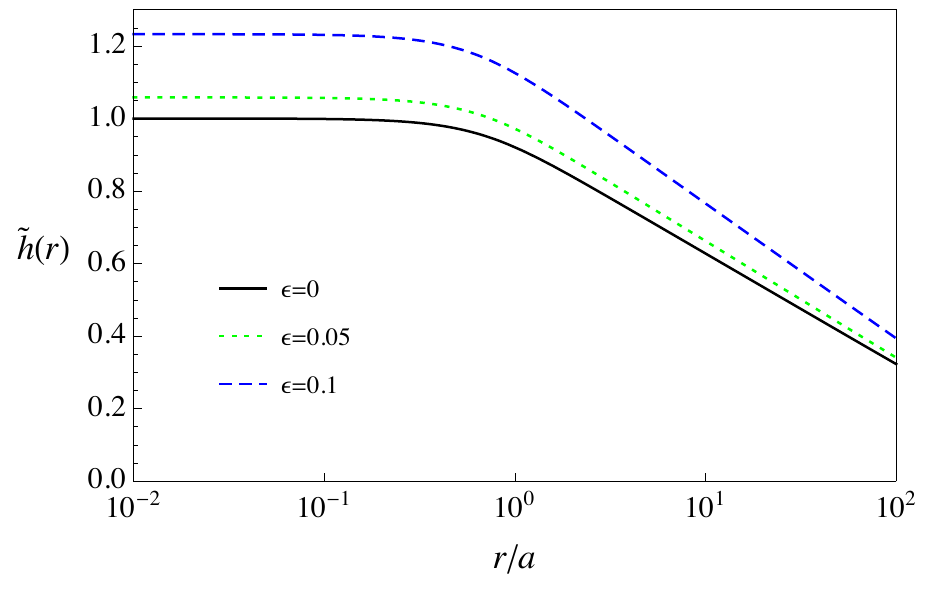}
\caption{Log-linear plot of the relative deformation profile $\tilde{h}(r)=h(r)/h^{(0)}(0)$, for a fixed probe-interface distance $d=\sqrt{2}a$
and $\mathcal{B}=10^{-3}$.}
\label{figh}
\end{figure}

Eq.~(\ref{younglapl}) 
is then solved using the method of matched asymptotic discussed earlier 
[see eq.~(\ref{matched})].
Within these assumptions, the bare solution reads
\begin{equation}
h^{(0)}(r) = \frac{\mathcal{A}}{4} \frac{a^5}{ H^4} 
 \left[
\text{K}_0 \left( \mathcal{B} \frac{r}{a} \right)   
 + \ln \left(\frac{r}{D} \right)  + \frac{H^2}{2D^2} \right]  \ ,
\end{equation}
where we define $D^2=d^2+r^2-a^2$ and $H^2=d^2-a^2$.
This result was previously derived in~\cite{Qui13}. The new contribution
is the fluctuation-induced correction to the bare profile. It  is given by
\begin{align}
h^{(1)}(r) = \frac{\mathcal{A}}{2}  \frac{a^7\Delta^2}{ H^8} 
& \Big[
\text{K}_0 \left( \mathcal{B} \frac{r}{a} \right) 
 + \ln \left(\frac{r}{D} \right)  
   \nonumber \\
 & + \frac{H^2}{2D^2} +\frac{H^4}{4D^4} + \frac{d^2H^6}{\Delta^2D^6}
  \Big]  \ ,
\end{align}
with $\Delta^2 =  5d^2+a^2$. The full solution $h=h^{(0)}+\varepsilon^2 h^{(1)}$ is plotted in fig.~\ref{figh} for 
different values of $\varepsilon$, and for $d=\sqrt{2}a$.
To illustrate the discussion, we consider for instance the deformation at the $r=0$ induced by a particle located at $d=\sqrt{2}a$.
Given the fact the $\mathcal{B}\ll 1$, 
we then have 
\begin{equation}
h(0) \simeq - \frac{\mathcal{A}a}{4} \ln \mathcal{B} \left( 1 + 22 \varepsilon^2 \right)   \ .
\label{h0}
\end{equation}
For $a\approx10$~nm and $\xi_{\perp}\approx1$~nm, we find that the correction
 contributes to $\approx 20 \%$ of the total deformation at the origin. 
 Note that this proportion only depends of the distance $d$ between the probe and the interface.

\section{Discussion}

Eq.~(\ref{h0}) reveals that fluctuations cannot be ignored when probing nanoscale deformations of a liquid interface.
Such a large contribution  was \textit{a priori} quite unexpected, but it also decreases very rapidly with $\varepsilon$: it drops to $5 \%$ when $\varepsilon \approx 0.05$,
and is completely negligible when $\varepsilon \approx 0.01$. A fine knowledge of the interface MSD is therefore required in order to interpret experimental data. 

In the model, it is implicitly assumed that the position of the probe is fixed. 
But in a real AFM experiment, the tip behaves as an harmonic oscillator and is itself  subject to Brownian motion. Consider for instance
ref.~\cite{Led12a}: in this experiment,
the tip oscillates at frequency $f\approx15$~kHz with a MSD $\langle z^2 \rangle \approx 10$~nm$^2$. The question is then to know whether fluctuations of the tip are  dynamically coupled to  interfacial fluctuations~\cite{Bic06,Bic07}. At the nanometer scale, interfacial modes are overdamped with relaxation rate 
$\gamma_q = \sigma q/  \eta$, with $q=2  \pi /\lambda$ the wave number and
 $\eta$ the mean viscosity of the two fluids. Taking $\sigma \approx 10^{-1}$~N/m and $\eta\approx10^{-3}$~Pa.s, we find that $\gamma_q \approx 10^{10}$~s$^{-1}$ for a wavelength of the order of  the tip radius  $\lambda\approx a\approx10$~nm. 
 Since the relaxation rate of interfacial fluctuations is several orders of magnitude higher than the frequency of the tip, the stationary approach is then fully justified.  

Still, dynamical issues are expected to be relevant in systems with a very low surface tension, for instance in near-critical fluids~\cite{Aar04}.
More generally, the viewpoint of dynamical coupling between tip and interface fluctuations is compelling since
it would allow to directly relate  liquid properties (surface tension, viscosity) to the statistical properties of the probe, with no particular assumption regarding interfacial
deformation.
Work on this issue is currently under progress.

\acknowledgments
The author wishes to thank D.~Dean and J.~Indekeu for insightful discussions.

\end{document}